\documentclass[11pt,a4paper]{article} % JHEP
\usepackage{jheppub}                  % JHEP
\usepackage{graphicx}
\pdfoutput=1
\usepackage{lineno}
\usepackage{cancel}
\usepackage{here} 

\usepackage[normalem]{ulem}	 % Part of the standard distribution

\def\lsim{\raise0.3ex\hbox{$\;<$\kern-0.75em\raise-1.1ex
\hbox{$\sim\;$}}}
\def\gsim{\raise0.3ex\hbox{$\;>$\kern-0.75em\raise-1.1ex
\hbox{$\sim\;$}}}

\title{ 
Constraint on Neutrino Decay with Medium-Baseline 
Reactor Neutrino  Oscillation Experiments 
}
\author{Thamys~Abrah\~ao$^{1}$,}
\author{Hisakazu Minakata$^{2,3}$,}
\author{Hiroshi Nunokawa$^{1,4}$,}
\author{and Alexander  A. Quiroga$^{1}$}
\affiliation{
$^1$Departamento de F\'{\i}sica, Pontif{\'\i}cia Universidade Cat{\'o}lica 
do Rio de Janeiro, C. P. 38071, 22452-970, Rio de Janeiro, Brazil  \\
$^2$Instituto de F\'{\i}sica, Universidade de S\~ao Paulo, C.\ P.\
66.318, 05315-970 S\~ao Paulo, Brazil \\
$^3$Instituto de F\'{\i}sica Te\'orica, UAM/CSIC, Calle Nicol\'as Cabrera 13-15, Cantoblanco E-28049 Madrid, Spain \\
$^4$Institute for Nuclear Theory, University of Washington, Box 351550, 
Seattle, WA 98195-1550, USA \\
\\ 
}    %
\abstract{ 
The experimental bound on lifetime of $\nu_3$, the neutrino mass
eigenstate with the smallest $\nu_{e}$ component, is much weaker than 
those of $\nu_1$ and $\nu_2$ by many orders of magnitude to which 
the astrophysical constraints apply. 
We argue that the future reactor neutrino oscillation experiments with 
medium-baseline ($\sim 50$ km), such as JUNO or RENO-50, has the best 
chance of placing the most stringent constraint on $\nu_3$ lifetime 
among all neutrino experiments which utilize the artificial source neutrinos. 
Assuming decay into invisible states, we show by a detailed $\chi^2$ 
analysis that the $\nu_3$ lifetime divided by its mass, $\tau_3 / m_3$, 
can be constrained to be $\tau_3/m_3 > 7.5 \ (5.5)\ \times 10^{-11}$
s/eV at 95\% (99\%) C.L. by 100 kt$\cdot$years exposure by JUNO. 
It may be further improved to the level comparable to 
the atmospheric neutrino bound by its longer run. 
We also discuss to what extent $\nu_3$ decay affects mass-ordering 
determination and precision measurements of the mixing parameters.  
}

\emailAdd{thamys.abrahao@fis.puc-rio.br}
\emailAdd{minakata@fmail.if.usp.br}
\emailAdd{nunokawa@puc-rio.br}
\emailAdd{alarquis@fis.puc-rio.br}

\begin{document} % JHEP 
%\pagewiselinenumbers
%\linenumbers

\begin{flushright}
INT-PUB-15-024
\end{flushright}
\hfill

\maketitle
\flushbottom

\section{Introduction}
\label{sec:introduction}

Investigations on the possibility of neutrino decay has a long history,
see e.g., \cite{Cowsik:1977vz,De Rujula:1980qd}. 
Since neutrino radiative decay is so tightly 
constrained~\cite{Agashe:2014kda}, decay into invisible final states 
are more commonly discussed, for example, in the context of majoron 
models~\cite{Chikashige:1980ui,Gelmini:1980re}.\footnote{
To be testable by various means we mention here, one has to arrange the majoron models such that neutrino lifetime is not too long \cite{Schechter:1981cv}.}
The bound on neutrino lifetime $\tau$ depend on (1) whether 
daughter neutrinos are active or sterile \cite{Barger:1999bg}, 
and in the former case (2) which neutrino mass ordering, normal or
inverted, is realized. 
It also depends on (3) whether the neutrinos are Dirac or Majorana
particles. However, in any one of these cases whenever the bound exists,
its order of magnitude is given by the condition ${ \tau }/{ m } \sim {
L }/{ E }$ for neutrinos with mass $m$ and energy $E$ that traverses
distance $L$ \cite{Beacom:2002cb}. 
We refer the condition as the {\em kinematic} estimate.  
It is nothing but stating that decay effect is sizeable when traveling 
time $t$ is comparable to lifetime $\tau_{ \text{lab} }$ 
of neutrinos in the laboratory (i.e., observer's) frame. 

Astrophysical neutrinos, because of their long path lengths, 
are the promising sources for yielding the stringent bounds on neutrino lifetime. 
The solar neutrinos have been used to place bound on the lifetime of $\nu_2$, which is the dominant component of $^8$B neutrinos under the assumption that $\nu_2$ decays into $\nu_1$ ($\bar{\nu}_1$) or into sterile states~\cite{Beacom:2002cb,Joshipura:2002fb,Bandyopadhyay:2002qg,Berryman:2014qha,Picoreti:2015ika}. In most cases the daughter neutrinos were assumed to be unobserved even if the decay into active neutrino is considered. 
The obtained bound is of the order of 
the kinematic estimate, $\tau_2/m_2 \sim 10^{-4}\ { \text{s} }/{ \text{eV} }$. 
We note that the authors of ref.~\cite{Berryman:2014qha} argued that it is 
possible to constrain also the lifetime of $\nu_1$ by considering 
low energy $pp$ and $^7$Be solar neutrinos, 
which have large $\nu_1$ components. 
Because of lower neutrino energies, the obtained bound for $\nu_1$ 
is better by about a factor of five than that for $\nu_2$~\cite{Berryman:2014qha}. 

Supernova neutrinos are potentially the most powerful source for bound on neutrino decay \cite{Frieman:1987as} which could lead to the bound ${ \tau }/{ m } \sim 10^{5}\ { \text{s} }/{ \text{eV} }$ according to the kinematic estimate.  
However, the available data is currently limited to the one which came from SN1987A~\cite{Hirata:1987hu,Bionta:1987qt}. 
Moreover, the bound applies only to $\nu_1$ and/or $\nu_2$ which have
large $\bar{\nu}_{e}$ components. Astrophysical neutrinos which have
been observed by IceCube \cite{Aartsen:2013jdh}, in principle, are
equally (or more) powerful as supernova neutrinos. 
But, to place the bound on ${ \tau }/{ m }$ we need either
identification of the sources, or complete determination of 
the neutrino flavor ratios~\cite{Beacom:2002vi}. 
Its typical order is estimated as ${ \tau }/{ m } \sim 10^{4}\ { \text{s} }/{ \text{eV} }$ assuming 1 TeV neutrinos from AGN at 100 Mpc \cite{Beacom:2002cb}. See \cite{Baerwald:2012kc} for recent discussions on the hypothesis of neutrino decays over cosmological distances. 

Leptonic decay of mesons leads to a bound on the majoron coupling constant with neutrinos $g_{\alpha \beta}$ $(\alpha, \beta=e, \mu, \tau)$ \cite{Barger:1981vd} of the order of $g^2 \sim 10^{-5}-10^{-4}$. For a comprehensive treatment of the majoron coupling bound with the pseudo-scalar as well as the scalar couplings, see e.g., \cite{Lessa:2007up} and an update \cite{Farzan:2011tz}. 
When translated into $\nu_2$ lifetime assuming decay into active
$\nu_1$, it gives a bound on $\tau_2/m_2$ in a range comparable with 
to somewhat weaker than the solar neutrino bound \cite{Beacom:2002cb}. 
Using lepton decay channels the authors of \cite{Lessa:2007up} also obtained the bound of couplings which include $\nu_\tau$, 
$\sum_{\alpha = e, \mu, \tau} \vert g_{\tau \alpha} \vert^2 < 0.1$, which would lead to a less stringent  bound on $\tau_3/m_3$ for active $\nu_3$ decay.

Therefore, so far no stringent bound on $\nu_3$ lifetime appears to
exist either from astrophysical or from laboratory neutrinos. 
Probably, the best way to constrain $\nu_3$ lifetime would be to use neutrino
oscillation phenomenon. It is because in this case one can select the
channels or region of kinematical phases that are sensitive to $\nu_{3}$
decay effect. 
Then, it is natural to use 
the oscillation driven by $\Delta m^2_{32} \simeq \Delta m^2_{31}$,
hereafter referred to as atmospheric-scale neutrino oscillation for simplicity. We will see that despite quantum mechanical nature of the phenomenon the kinematic estimate applies. The bound on $\nu_3$ lifetime was obtained \cite{GonzalezGarcia:2008ru} by using the data collected by the Super-Kamiokande (SK) atmospheric neutrino observation~\cite{Wendell:2010md} as well as by the long-baseline accelerator experiments. 
See \cite{Gomes:2014yua} for a recent update of the accelerator bound. 

In this paper, under the assumption of decay into invisible daughters,
we analyze the bound on $\nu_3$ lifetime which will be placed by the
future medium-baseline reactor neutrino experiments such as JUNO
\cite{He:2014zwa} or RENO-50 \cite{Park:2014sja} 
via observing the distortion of neutrino energy spectrum. 
In fact, we argue that they could provide, in principle, 
the most stringent bound on $\nu_3$ lifetime among all 
the artificial source neutrino experiments. It is because they can observe the effect of atmospheric-scale oscillations at the baseline around the maximum of the solar-scale oscillation, $L_\text{OM} =4\pi E/\Delta m^2_{21}$, which is longer than $L_\text{OM}$ for the atmospheric-scale oscillation by a factor of $\simeq 30$. Assuming five years operation of JUNO, we obtain the bound ${ \tau_{3} }/{ m_{3} } \gsim 7.5\ (5.5) \times10^{-11}$ s/eV at 95\% (99\%) CL.
See section~\ref{sec:uniqueness} for details, and for 
the relationship between our argument and 
the bound obtainable by using atmospheric neutrino data. 

 The principal objective of the medium-baseline reactor neutrino experiments is to determine the neutrino mass ordering. Then, the immediate question is whether it would be disturbed if possibility of neutrino decay is taken into account in fitting the data, or more drastically, when $\nu_3$ would actually decay. 
It was also noticed that such experiment has a potential of determining
the mixing parameters in a high precision
\cite{Minakata:2004jt,Bandyopadhyay:2004cp}. In fact, the recent works
with much more elaborate treatment of experimental errors reported an
extreme precision of sub-percent level for $\sin^2 \theta_{12}$ and
$\Delta m^2_{21}$ \cite{Ge:2012wj,Capozzi:2013psa}. 
Then, the natural question is whether or to what extent these sensitivities could be affected when possibility of $\nu_3$ decay is turned on. In section~\ref{sec:impact-decay} we address these questions.

\section{Uniqueness of the medium-baseline reactor neutrino experiments}
\label{sec:uniqueness}

In this section we try to convince the readers that JUNO/RENO-50 is, in principle, the highest sensitivity experiment among all those which utilize artificial source (or beam) neutrinos in detecting the possible decay effect of $\nu_3$. 
In this paper, 
we consider the case of invisible decay of $\nu_3$. 
That is, we assume that the decay products are either some sterile
states, or can involve active $\nu_1$ and/or $\nu_2$ state but with 
significant energy degradation such that daughter neutrinos cannot be
observed. 
The latter possibility necessitates the neutrino mass ordering to be the normal type. 

When $i$-th mass eigenstate neutrino decays with lifetime $\tau_{i}$ at
rest, the energy $E_{i}$ (more precisely, the energy difference
normalized appropriately) 
of propagating $i$-th mass eigenstate neutrino can be written as 
\begin{eqnarray}
E_{i} = \frac{ m_{i}^2 }{2E} - i \frac{ \Gamma_{i} }{2} 
\label{ith-energy}
\end{eqnarray}
where 
\begin{eqnarray}
\frac{1}{ \Gamma_{i} }= \left( \frac{E}{m_{i}} \right) \tau_{i} 
\label{width}
\end{eqnarray}
is a Lorentz dilated lifetime. 

As will be shown in appendix~\ref{sec:probabilities} 
the decay of $\nu_3$ produces the following two characteristic
modifications in the oscillation probabilities in vacuum:\footnote{
%%%%%%%%%%%%% footnote %%%%%%%%%%%%%%
The matter effect does not appear to affect 
the conclusion of this section in a significant way. 
}
\begin{itemize}

\item 
Reduction of the atmospheric-scale oscillation amplitude with the form 
$\cos \left( \frac{ \Delta m^2_{\text{atm}} L}{2 E} \right) e^{- \frac{ \Gamma_{3} L}{2} }$. 

\item 
Decrease of normalization of the probabilities by the amount
proportional to $\left( 1 - e^{- \Gamma_{3} L } \right)$. 

\end{itemize}
Here, $L$ is the distance traveled by neutrinos. To make the point clearer we have used an approximation $\Delta m^2_{31} \approx \Delta m^2_{32} \equiv \Delta m^2_\text{atm}$ where $\Delta m^2_{ji} \equiv m^2_{j} - m^2_{i}$. 
The first feature stems from the fact that the effect comes from the interference between the first$-$second mass eigenstates and the third. The second feature represents the effect of decaying $i$-th mass eigenstate projected onto the initial and final neutrino flavor states. It is independent of neutrino oscillation and exists even in the limit of vanishing $\Delta m^2_{ji}$, as it must. This effect is prominent in $\nu_{\mu}$ (and $\bar{\nu}_\mu$) disappearance channel, but is negligible for $\nu_e$ (and $\bar{\nu}_e$) survival probability. Whereas the first effect is equally important in the all oscillation channels, as will be shown in appendix~\ref{sec:probabilities}. 

The decay effect is negligible for baseline $L$ with $\Gamma_{3} L \ll
1$, and it becomes significant only when $\Gamma_{3} L \gsim 1$. 
In most of the 
oscillation experiments which use man-made neutrino sources, 
the baseline is set to the first oscillation maximum of
atmospheric-scale oscillation, ${ \Delta m^2_\text{atm} L}/{(2 E)}
\simeq \pi$. In such setting, it is likely 
that we obtain the lower bound on the width $\Gamma_{3}$ as
\begin{eqnarray}
\frac{1}{ \Gamma_{3} } \lsim L \simeq \frac{2 \pi E} { \Delta m^2_\text{atm} }.
\label{Gamma-bound}
\end{eqnarray}
From (\ref{width}), we know that 
$\tau_{3}/m_{3}= (\Gamma_{3} E)^{-1}$.
Using this relation in (\ref{Gamma-bound}) 
one can estimate the achievable upper bound on $\tau_{3}$ in such setting as 
\begin{eqnarray}
\frac{ \tau_{3} }{ m_{3} } \simeq \frac{2 \pi } { \Delta m^2_\text{atm} } 
\hspace{3mm}
\left( = \frac{L}{E} \right), 
\label{tau3-bound1}
\end{eqnarray}
where the last equality in parenthesis assumes that the distance $L$ is
at the first oscillation maximum. 
It implies that the kinematic estimate applies. Using (\ref{tau3-bound1}), one can estimate the upper bound on $\tau_{3} /m_{3}$ at 
\begin{eqnarray}
\frac{ \tau_{3} }{ m_{3} } & \simeq & 
1.7 \times 10^{-12} 
\left( \frac{ \Delta m^2_\text{atm} }
{ 2.4 \times10^{-3}\ \text{eV}^2 } \right)^{-1} 
\frac{\text{s}}{\text{eV}}.
\label{tau3-bound2}
\end{eqnarray}

In the JUNO or RENO-50 setting, 
hereafter referred simply as JUNO setting for simplicity,
the detector is placed approximately at the maximum of the solar-scale oscillation, 
${ \Delta m^2_{21} L}/{(2 E)} \simeq \pi$. 
Then, we can achieve about 30 times longer lifetime bound
\begin{eqnarray}
\frac{ \tau_{3} }{ m_{3} } &\simeq& 
5.5 \times 10^{-11} 
\left( 
\frac{ \Delta m^2_{21} }{ 7.5 \times10^{-5}\ \text{eV}^2 } 
\right)^{-1}  
\frac{\text{s}}{\text{eV}}.
\label{tau3-bound2}
\end{eqnarray}
It should be stressed that the JUNO setting is unique 
(among artificial neutrino source experiments) in making observation of 
the atmospheric-scale neutrino oscillation at the distance of 
the solar-scale oscillation maximum possible.  
This completes our argument that JUNO is 
the highest sensitivity experiment among (practically) all 
the ongoing or proposed artificial-source neutrino experiments as far as 
the $\nu_3$ lifetime bound is concerned. 

Of course, the above argument does {\em not} necessarily imply that 
the JUNO bound on $\nu_3$ lifetime must be the severest one achievable
by all the neutrino experiments.  
The likely (and probably unique) exception is the one placed by observation of the atmospheric neutrinos. The naive kinematic estimate for the $\nu_3$ lifetime would lead to the sensitivity to 
\begin{eqnarray}
\frac{ \tau_{3} }{ m_{3} } 
& \simeq & \frac{L}{E} 
\approx \frac{ 10^{4}~\text{km} }{ 1~\text{GeV} }
\simeq 
3.3 \times 10^{-11} 
\frac{\text{s}}{\text{eV}}, 
\label{tau3-bound-atm}
\end{eqnarray}
which suggests that the bound by the atmospheric neutrinos is comparable
as the one by JUNO. However, in the case of atmospheric neutrino
experiments they can observe neutrinos in much wider energy and baseline
ranges than those used in the above estimate.  
Therefore, we expect that the bound on $\tau_{3} / m_{3} $ from the
atmospheric neutrinos is tighter than the naive estimate given in
(\ref{tau3-bound-atm}). 
In fact, the authors of ref.~\cite{GonzalezGarcia:2008ru}, by using 
the SK atmospheric neutrino data and the others, obtained 
the bound $\tau_{3}/m_{3} > 9.3 \times 10^{-11}$ s/eV at 99\% CL, 
which is stronger than the one shown in Eq. (\ref{tau3-bound-atm}) 
by about a factor of 3.\footnote{
%%%%%%%%%%%%%% footnote %%%%%%%%%%%%%%%
For comparison we note that the kinematic estimate for JUNO,
(\ref{tau3-bound2}), is different from our results based on $\chi^2$
analysis by $\simeq 40\%$ only. It should also be noticed that if a
re-analysis in \cite{GonzalezGarcia:2008ru} would be done with the
currently accumulated SK data, the bound should become even tighter. 
%%%%%%%%%%%%%% footnote %%%%%%%%%%%%%%%
%
It is also more stringent by a factor of 1.7 than our JUNO five-years bound to be obtained in section~\ref{sec:results-bound}. 
}

\section{Effect of $\nu_3$ decay on the oscillation probabilities and the observable}
\label{sec:prob-event}

The $\bar{\nu}_e$ survival probability relevant for reactor neutrinos
for the baseline $L$ in vacuum is given, under the approximation 
$\Delta m^2_{31} \approx \Delta m^2_{32} \equiv \Delta m^2_{\text{atm}}$, as
\begin{eqnarray}
P(\bar{\nu}_e \rightarrow \bar{\nu}_e) 
&=& 
1 -  
c^4_{13} \sin^2 2\theta_{12} 
\sin^2 \left( \frac{\Delta m_{21}^2 L}{4E} \right) 
\nonumber \\ 
&-&
s^4_{13} \left( 1 - e^{- \Gamma_{3} L } \right) 
- \frac{1}{2} \sin^2 2\theta_{13} 
\left[ 1 - 
\cos \left( \frac{\Delta m^2_\text{atm} L}{2 E} \right) 
e^{- \frac{\Gamma_{3} L}{2} } 
\right],
\label{P-ee1}
\end{eqnarray}
where $c_{ij} \equiv \cos \theta_{ij}$ and $s_{ij} \equiv \sin \theta_{ij}$. See appendix~\ref{sec:probabilities} for derivation. 
For simplicity and as a good approximation, we ignore the matter effect in this work. 
As we stated in the previous section, there are two types
of terms which are affected by the neutrino decay. 
However, they come with vastly different magnitudes in 
the $\bar{\nu}_e$ channel; 
since the current neutrino data implies $s^2_{13} \simeq 0.02$, 
the coefficients of $\left( 1 - e^{- \Gamma_{3} L } \right)$ 
and $\cos \left( \frac{\Delta m^2_\text{atm} L}{2 E} \right)$ 
terms are, respectively, $s^4_{13} \sim 5 \times 10^{-4}$ and 
$\frac{1}{2} \sin^2 2\theta_{13} \sim 4 \times 10^{-2}$.
Therefore, the former oscillation-independent 
$\bar{\nu}_{e}$ attenuation effect should be negligible.

%
%%%%%%%%%%%%%%%%%%%%%%%%%%%%% FIG 1 %%%%%%%%%%%%%%%%%%%%%%%%%%%%%%
\begin{figure}[t!]
\begin{center}
\vspace{-0.2cm}
\hspace{0.5cm}
\includegraphics[bb=0 0 792 612,width=0.9\textwidth]{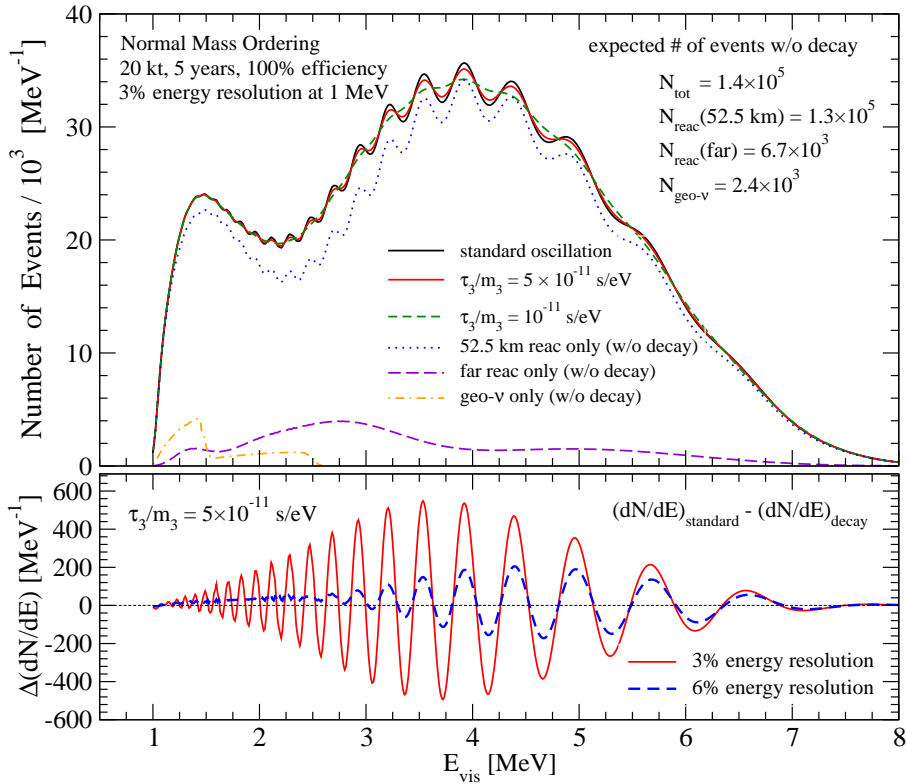}
\end{center}
\vspace{-0.8cm}
\caption{
Upper panel shows the expected event distribution as a function 
of the visible energy of positron at the 20 kt detector placed at $L=52.5$ km from 
the reactor complex of 35.8 GW thermal power. 
The case without decay effect is indicated by 
the black solid curve whereas the case with decay effect
is shown by the red solid and green dashed
curves, corresponding, respectively, 
to $\tau_3/m_3 = 5 \times 10^{-11}$ and $10^{-11}$ s/eV. 
We also show the individual contributions coming from 
the reactor complex at the medium-baseline $L=52.5$ km (blue dotted curve), 
from the far reactor complexes located at Daya Bay with $L$=215 km and 
Huizhou with $L$ = 265 km (violet dashed curve)
and geoneutrinos (orange dash-dotted curve). 
In the lower panel, in order to see clearly 
the importance of the energy resolution, 
we show, for $\tau_3/m_3 = 5 \times 10^{-11}$ s/eV, 
the difference of the cases without and with decay effect, 
or $dN/dE$ (without decay) - $dN/dE$ (with decay), 
as a function of $E_\text{vis}$
for 3\% and 6\% energy resolution, respectively, 
by the red solid and blue dashed curves. 
}
\label{fig:Spectrum1}
\end{figure}
%%%%%%%%%%%%%%%%%%%%%%%%%%%% FIG 1 %%%%%%%%%%%%%%%%%%%%%%%%%%%%%%

Then, the question is how such decay-affected oscillation probability
manifests itself into the observable quantities. 
To give a feeling to the readers we show in the upper panel of 
Fig.~\ref{fig:Spectrum1} the energy spectrum of events as 
a function of positron deposited energy, or visible energy $E_\text{vis}$,\footnote{
%%%%%%%%%%%%% footnote %%%%%%%%%%%%%
The visible energy $E_\text{vis}$ is approximately related 
to neutrino energy $E$ 
as $E_\text{vis} \simeq E - (m_n-m_p) + m_e$, 
where $m_n$, $m_p$, $m_e$, are, respectively 
the mass of neutron, proton and electron. 
}
calculated by convoluting the neutrino flux and 
the cross section of the inverse $\beta$-decay (IBD) ($\bar{\nu}_e + p
\to e^+ + n$) reaction. 
The three curves with different combinations of colors and line types 
correspond to the following three cases: 
no decay (black solid line), $\tau_3/m_3 = 5 \times 10^{-11}$ s/eV (red solid line), and $\tau_3/m_3 = 10^{-11}$ s/eV (green dashed line). 
See appendix \ref{sec:spectrum} for the procedure to obtain these curves. 
%%%%%%%% moved to here %%%%%%%%%
As we can see from the upper panel in Fig.~\ref{fig:Spectrum1}, 
the decay of $\nu_3$ state tends to average out the fast oscillation 
(wiggles) driven by $\Delta m^2_\text{atm}$.\footnote{
%%%%%%%%%%%%% footnote %%%%%%%%%%%%%
In this connection we note that quantum decoherence \cite{Lisi:2000zt} has the similar effect on energy spectrum of $\bar{\nu}_{e}$, and the relevant phenomenology is discussed with the JUNO setting in ref.~\cite{Bakhti:2015dca}. 
}

Since decreasing $\theta_{13}$ and $\nu_3$ decay both reduce 
the amplitude of atmospheric-scale oscillation, they could potentially 
be confused with each other.  
Fortunately, the confusion is precluded by the precision measurement 
of $\theta_{13}$ done by the short baseline reactor neutrino
experiments. 
(See the  related comment at the end of section~\ref{sec:results-bound}.) 
Currently, $\sin^2 2\theta_{13}$ is measured with the uncertainty of 
$\simeq$ 6\% \cite{An:2015rpe}, 
while a possibility of reaching the ultimate error of $\simeq 3\%$ by 
the end of 2017 is mentioned in \cite{DayaBay-Nutele}.

Degradation of the oscillation amplitude can also occur by 
a finite energy resolution of the detector, which would cause another 
confusion with the decay effect. To illustrate this point, we show in 
the lower panel of Fig.~\ref{fig:Spectrum1}, the difference between 
the energy spectra in the absence and in the presence of $\nu_3$ decay 
for two cases of energy resolution, 3\% (red solid curve) 
and 6\% (blue dashed curve) at 1 MeV. 
Roughly speaking, doubling the energy resolution causes an amplitude attenuation of the signal by a factor of three. We observe that the detectability of the decay effect is strongly dependent on the energy resolution. Fortunately, the 3\% energy resolution is to be reached by JUNO and RENO-50~\cite{He:2014zwa,Park:2014sja}.

To our understanding, the remaining  questions which need to 
be addressed are as follows: 
\begin{itemize}
\item 
As shown in Fig.~\ref{fig:Spectrum1}, the decay effect is to reduce 
the atmospheric-scale oscillation amplitude. 
This is {\em the oscillation} to be utilized to determine the mass ordering in JUNO. Then, the obvious question is to what extent the mass ordering determination could be disturbed if the effect of $\nu_3$ decay is taken into account. 

\item 
Another relevant question would be to what extent the sensitivities to
mixing parameter measurement in JUNO could be disturbed
by $\nu_3$ decay. 

\end{itemize}
\noindent
We will discuss these issues in section~\ref{sec:impact-decay} from 
the following two different viewpoints (assumptions): 
(1) There is no decay in the input data set, 
but we consider the decay effect in the output (fit), 
and 
(2) $\nu_3$ actually decays with the lifetime which is marginally 
consistent with the current and the expected JUNO bound. 

\section{Analysis method}
\label{sec:method}

For definiteness, throughout the paper (including Fig.~\ref{fig:Spectrum1}), we assume that the true (input) values of the oscillation parameters are given as follows, 
\begin{eqnarray}
\Delta m^2_{21} &=& 7.50\times 10^{-5} \text{eV}^2, \ \ 
\sin^2 \theta_{12} = 0.304, \nonumber \\ 
\Delta m^2_{31} &=& 2.46\times 10^{-3} \text{eV}^2,  \ \ 
\sin^2 \theta_{13} = 0.0218, 
\label{eq:input-osc}
\end{eqnarray}
which are taken from the best fitted values of the one of
the recent global analysis~\cite{Gonzalez-Garcia:2014bfa}, 
and assume the normal mass ordering ($\Delta m^2_{31} > 0$),
unless otherwise stated.

In our statistical analysis, we formally divide the $\chi^2$ function 
in three terms as 
\begin{eqnarray}
\chi^2 \equiv 
\chi^2_\text{stat} 
+\chi^2_\text{param} 
+\chi^2_\text{sys}, 
\label{eq:chi2}
\end{eqnarray}
as done in \cite{Ge:2012wj,Capozzi:2013psa}.
The first term, $\chi^2_\text{stat}$, 
is computed by taking the limit of 
infinite number of bins 
which is justified because of the large 
number of events expected at 20 kt detector, 
$\simeq 1.4 \times 10^5$ for 
5 years of operation, 
as follows~\cite{Ge:2012wj,Capozzi:2013psa},
\begin{eqnarray}
\chi^2_\text{stat} \equiv 
\int_0^{{E^\text{max}_\text{vis}}} 
dE_\text{vis}
\left(
\frac{
\displaystyle
 \frac{dN^\text{obs}}{dE_\text{vis}}
-\sum_{i=\text{reac, U, Th}}
(1+\xi_i)\frac{dN^\text{fit}_i}{dE_\text{vis}} }
{\displaystyle
\sqrt{ 
\frac{dN^\text{obs}}{ dE_\text{vis}} 
} 
}
\right)^2,
\label{eq:chi2_stat}
\end{eqnarray}
where 
$dN^\text{obs} / dE_\text{vis}$
is the event distributions of the observed (simulated) signal, 
and $\xi_i$ is the flux normalization parameters for reactor neutrinos as well as for geoneutrinos to be varied freely subject to the pull term in $\chi^2_\text{sys}$ (see below) and we integrate up to $E^\text{max}_\text{vis} = 8$ MeV. 
In our analysis, following \cite{Capozzi:2013psa}, we include the
contributions not only from Yangjiang and Taishan reactor complexes  at
$L=52.5$ km (approximated by a single reactor with the thermal power of
35.8 GW) but also the contributions coming from the reactor complexes
located at Daya Bay ($L=215$ km) and Huizhou ($L=265$ km), as well as
geoneutrinos. See appendix \ref{sec:spectrum} for details.

The second term in (\ref{eq:chi2}) 
takes into account the current uncertainties of 
the standard oscillation parameters and 
is given by 
\begin{eqnarray}
\chi^2_\text{param} 
\equiv \sum_{i=1}^4 
\left( \frac{{\bar{x}}_i-x_i^{\text{fit}}}
{\sigma({x_i})} \right)^2,
\label{eq:chi2_param}
\end{eqnarray}
where $\bar{x}_i$ and $x_i^{ \text{fit} } $ 
($i=1$-4) denote, respectively, 
the assumed true (input) and fitted values
with 
$x_1 \equiv \sin^2\theta_{12}$, $x_2 \equiv \Delta m^2_{21}$, 
$x_3 \equiv \sin^2\theta_{13}$, $x_4 \equiv \Delta m^2_{31}$.
For the values of $\sigma(x_i)$, we take the current 1 sigma uncertainties determined by the global fit~\cite{Gonzalez-Garcia:2014bfa}, 
$\sigma(\sin^2\theta_{12}) = 4.1$\%, 
$\sigma(\Delta m^2_{21}) = 2.4$\%,
$\sigma(\sin^2\theta_{13}) = 4.6$\% 
and $\sigma(\Delta m^2_{31}) = 1.9$\%. 

The last term in (\ref{eq:chi2}), $\chi^2_\text{sys}$, 
takes into account the contributions of two kind of experimental 
systematic uncertainties we consider
\begin{eqnarray}
\chi^2_\text{sys} 
\equiv 
\left(\frac{\xi_\text{reac}^\text{fit} }
{ \sigma_{{\xi}_\text{reac}} } \right)^2 
+
\left(\frac{\xi_\text{U}^\text{fit} }
{ \sigma_{{\xi}_\text{U} }} \right)^2 
+
\left(\frac{\xi_\text{Th}^\text{fit} }
{ \sigma_{{\xi}_\text{Th} }} \right)^2 
+ 
\left(\frac{\eta^\text{fit} }
{ \sigma_{\eta} } \right)^2
\label{eq:chi2_sys}
\end{eqnarray}
where $\sigma_{\xi_\text{reac}}$, 
$\sigma_{\xi_\text{U}}$ and $\sigma_{\xi_\text{Th}}$
describe, respectively, 
the normalization uncertainties for reactor neutrinos,
uranium and thorium induced geoneutrinos. 
For them we take $\sigma_{\xi_\text{reac}}$ = 
3\%~\cite{Mueller:2011nm,Huber:2011wv} for reactor neutrinos and 
$\sigma_{\xi_\text{U}} = \sigma_{\xi_\text{Th}} = 20$\% for 
geoneutrinos following ref.~\cite{Capozzi:2013psa}.   
In addition to the normalization uncertainties, we also consider the uncertainty of the energy resolution by including the last term in (\ref{eq:chi2_sys}) with $\sigma_\eta$ (see below).

For simplicity, we assume 100\% detection efficiency, $\epsilon_\text{det} = 1$. 
The effect of uncertainty in the detection efficiency 
(IBD selection, fiducial volume, etc) is thus neglected, 
since it is difficult to reliably estimate. 
However, those errors, being energy independent, should hardly 
affect the sensitivity to shape-modulating effects caused by the neutrino decay 
as well as by the differing mass orderings. 

With regard to the energy resolution, we only consider the stochastic term, i.e., $\sigma_{E}/{E} = 0.03(1+\eta)/\sqrt{ E/\text{MeV} }$ contribution where $\eta$ is the parameter accounting for the energy resolution variation.\footnote{
%%%%%%%%%%%%% footnote %%%%%%%%%%%%%
See appendix~\ref{sec:spectrum} for the Gaussian energy resolution 
implementation in our analysis.}
The non-stochastic term(s) are expected to be the most relevant, even
dominant, at energies $>3$ MeV. However, this is a complex experimental
matter impossible to anticipate at this stage, so its impact is simply
neglected here. 
Some deterioration of the sensitivity should be expected, if 
those terms were considered. 
Because of the illustrated dependence of the energy resolution on the decay
sensitivity, its uncertainty is taken into account by using  
a pull term with $\sigma_\eta$ = 10\%.\footnote{
%%%%%%%%%%%%% footnote %%%%%%%%%%%%%%%
Notice that the energy resolution uncertainty, 
usually estimated by using calibration source data, 
has never been a sensitive quantity for physics 
parameter determination so far, but it must be carefully 
considered by experimental collaborations 
in completion of the detector design. }
We feel it appropriate to consider no other systematic error on 
the energy scale at this stage.  

The $\chi^2$ is computed in the following way: 
In order to derive the expected bound on 
the $\nu_3$ decay lifetime, 
for our input data, we consider  
the JUNO setup to compute $dN^\text{obs}/dE_\text{vis}$ 
assuming no decay ($\tau_3/m_3=\infty$) using 
the oscillation parameters given in (\ref{eq:input-osc}). 
Then we try to fit such $dN^\text{obs}/dE_\text{vis}$ by minimizing 
the $\chi^2$ function varying freely $\theta_{12}$, $\Delta m^2_{21}$,
$\theta_{13}$, $\Delta m^2_{31}$, ${\xi_\text{reac}}$, ${\xi_\text{U}}$, ${\xi_\text{Th}}$, $\eta$, and $\tau_3/m_3$ 
subject to the pull terms in (\ref{eq:chi2_param}) and (\ref{eq:chi2_sys}). 
In this way we can estimate the sensitivity to $\tau_3/m_3$ and 
derive the bound on it, and at the same time, can determine the allowed regions of other parameters as well.

Using the $\chi^2$ function, we will determine the allowed range of 
$\tau_3/m_3$ by the condition 
\begin{equation}
\Delta \chi^2 \equiv \chi^2 - \chi^2_{{\text{min}}} 
= 2.71,\ 3.84\ \text{and}\ 6.63\ (1,\ 4\ \text{and} \ 9),
\end{equation}
at 90\%, 95\% and 99\% (1, 2 and 3 $\sigma$) CL, respectively, for one degree of freedom. For the case where we show the allowed regions in the plane spanned by any combination of two out of nine parameters, we use the condition $\Delta \chi^2 = 2.3, 6.18$ and 11.83, respectively, for 1, 2 and 3 $\sigma$ CL for two degree of freedom. 
We note that $\chi^2_\text{min}=0$ by construction because we do not 
take into account the statistical fluctuation in simulating the artificial data. 

\section{The bound on $\tau_3/m_3$}
\label{sec:results-bound}

In this section, we derive the bound on $\tau_3/m_3$ assuming that
$\nu_3$ decays into invisible states, whereas $\nu_1$ and $\nu_2$ are
stable. For definiteness, we assume the normal mass ordering. 
But, given the expression of $P(\bar{\nu}_e \rightarrow
\bar{\nu}_e)$ in (\ref{P-ee1}), we expect that our results are valid
also for the case of inverted mass ordering.

As our standard setup, we assume 5 years running of the JUNO detector
with the fiducial volume of 20 kt which is placed at $L=52.5$ km away
from an effective single reactor with the thermal power of 35.8 GW. 
We assume the detector's running with 100\% efficiency, $\epsilon_\text{det} = 1$. But, if $\epsilon_\text{det} < 1$ the running time must be scaled to 
[5/$\epsilon_\text{det}$] years to obtain the same results. 
Since JUNO may take data for a longer period, we also 
consider the case of exposure for 15 years. 
See appendix~\ref{sec:spectrum} for details of our calculation.

%%%%%%%%%%%%%%%%%%%%%%%%%%%%%%%% FIG 2 %%%%%%%%%%%%%%%%%%%%%%%%%%%%%%%%%%
\begin{figure}[h!]
\begin{center}
\vspace{-1.2cm}
\hspace{0.8cm}
\includegraphics[bb=0 0 792 600,width=0.91\textwidth]{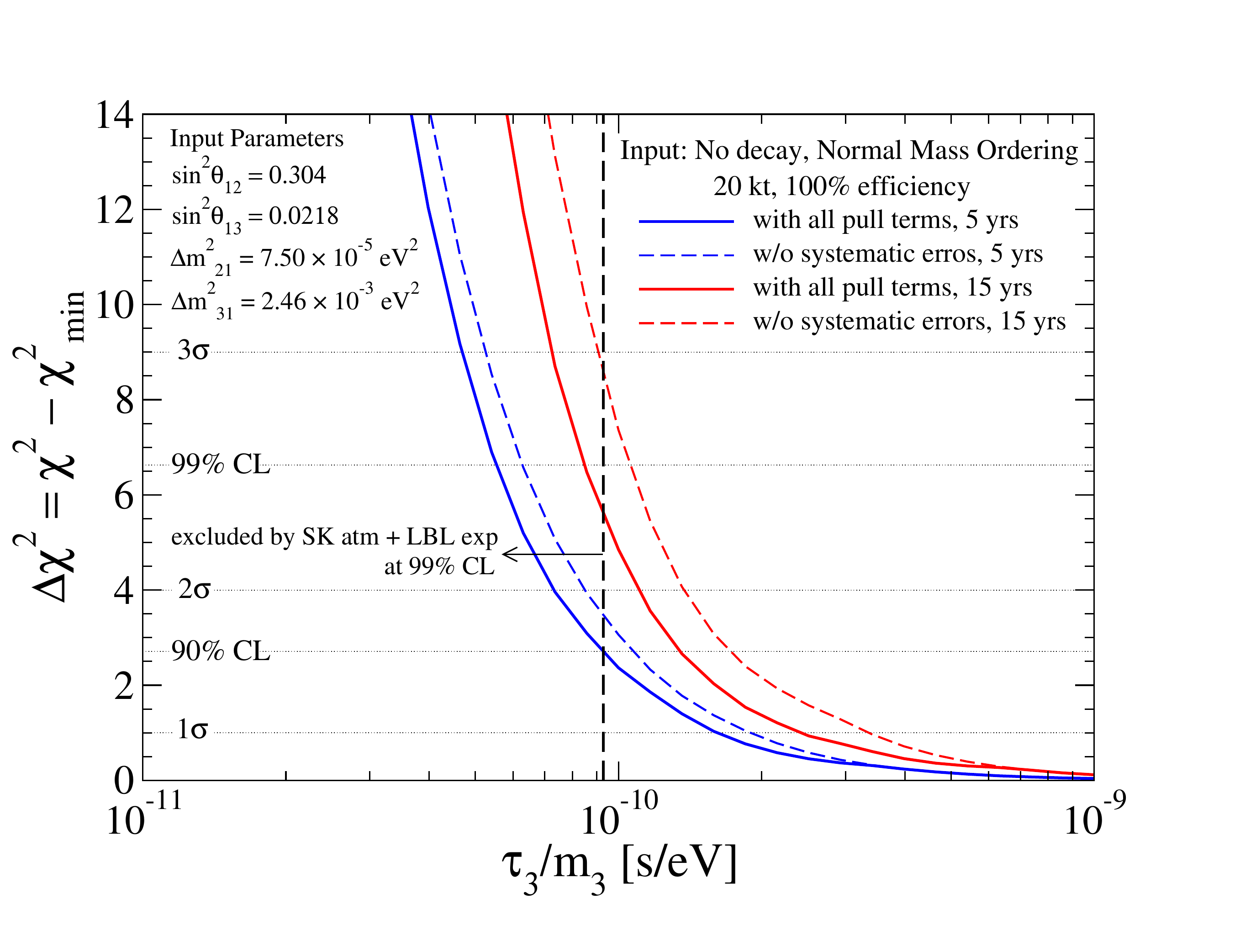}
\end{center}
\vspace{-1.2cm}
\caption{
$\Delta \chi^2 \equiv \chi^2-\chi^2_\text{min}$ 
is shown, by the red (blue) curves for 5 (15) years of data taking,  
as a function of the fitted value of 
$\tau_3/m_3$ calculated for the JUNO detector placed at 
$L=52.5$ km from a reactor with 35.8 GW thermal power, assuming 5 years
of exposure and 100\% detection efficiency.
We have taken that the true (input) value of $\tau_3/m_3$ is 
infinite (stable $\nu_3$). 
The solid curves correspond to the results obtained by 
using our full $\chi^2$ defined in (\ref{eq:chi2})
whereas the dashed ones correspond to the case without
assuming systematic errors. 
The contributions from the reactors at Daya Bay and Huizhou 
as well as those from geoneutrinos are taken into account. 
The bound comes from the SK atmospheric neutrinos plus 
long-baseline oscillation experiment obtained in \cite{GonzalezGarcia:2008ru} 
is also indicated by the vertical black dashed line. 
}
\label{fig:Delta_chi2}
\end{figure}
%%%%%%%%%%%%%%%%%%%%%%%%%%%%%%%% FIG 2 %%%%%%%%%%%%%%%%%%%%%%%%%%%%%%%%%%
%
In Fig.~\ref{fig:Delta_chi2} we show $\Delta \chi^2 \equiv
\chi^2-\chi^2_\text{min}$ as a function of the fitted value of
$\tau_3/m_3$ where the input (true) value of $\tau_3/m_3$ is assumed to
be infinity, i.e., $\nu_3$ is stable, with oscillation parameters given
in (\ref{eq:input-osc}). All the other eight parameters are marginalized
in the fit. 
The case of exposure for 5 (15) years is shown by the blue (red) solid curve. 
To exhibit the effect of the systematic error onto the bound, 
we also show by the dashed 
curves $\Delta \chi^2 $ for the case without all the systematic errors.

From our result shown by the solid blue curve in 
Fig.~\ref{fig:Delta_chi2}, we can conclude that if 
the data is consistent with no-decay hypothesis, the range 
\begin{eqnarray}
\frac{\tau_3}{m_3} < 7.5 \ (5.5)\ \times 10^{-11}\ 
\frac{\text{s}}{\text{eV}}
\label{eq:sensitivity-result}
\end{eqnarray}
can be excluded at 95 (99\%) \% CL by 5 years exposure by JUNO. 

As mentioned in section~\ref{sec:uniqueness}, this bound is about a
factor of 1.7 weaker than the bound $\tau_{3}/m_{3} > 9.3 \times
10^{-11}$ s/eV at 99\% CL based on the data of atmospheric neutrinos 
(as well as of long-baseline oscillation
experiments)~\cite{GonzalezGarcia:2008ru}. 
The bound is indicated by the vertical black dashed line in Fig.~\ref{fig:Delta_chi2}. 
By considering an extended running of 15 years, the JUNO bound 
can be improved to 11 (8.5) $\times 10^{-11}$ s/eV at 95 (99\%) \% CL, 
which is barely comparable to the atmospheric neutrino bound. 
A comparison between the bound of the order of 
$\tau_{3}/m_{3} \gsim 10^{-10}$ s/eV obtained here and 
in \cite{GonzalezGarcia:2008ru} to the one deduced by using 
lepton decay channel \cite{Lessa:2007up} is described 
in \cite{GonzalezGarcia:2008ru}. 
They obtained the bound on $\nu_3 -\nu_s-$majoron coupling as 
$\vert g_{s 3} \vert^2 \lsim 10^{-4} 
\left( m_s / 1\ \text{eV} \right)^{-2}$ at 90\% CL assuming $m_3 \gg m_s$,
where $m_s$ denotes the sterile neutrino mass.

We also considered the hypothetical situation where $\theta_{13}$ were
unknown 
at the time of JUNO running (not shown). We did it by removing the pull term for $\theta_{13}$ from our $\chi^2$, and found that the bound on $\tau_3/m_3$ would become worse by about a factor of three. It is because the $\nu_3$ decay and the effect coming from the uncertainty in $\theta_{13}$ can be confused with each other, as we mentioned in Sec.~\ref{sec:prob-event}.

\section{Impact of $\nu_3$ decay on the determination of the mass ordering and the oscillation parameters}  
\label{sec:impact-decay}

\subsection{Impact of $\nu_3$ decay on the mass ordering determination}
\label{sec:mass-ordering}

Now, we address the question of effect of $\nu_3$ decay onto the mass
ordering determination in JUNO/RENO-50, 
as promised at the end of
section~\ref{sec:prob-event}. We use the same analysis tool as used to
obtain the bound on $\nu_3$ lifetime in the previous section. The input mass ordering is always taken to be normal in this section. 

%%%%%%%%%%%%%%%% FIG 3 %%%%%%%%%%%%%%%%%%
\begin{figure}[h!]
\begin{center}
\vspace{-1.1cm}
\hspace{0.4cm}
\includegraphics[bb=0 0 792 612,width=0.80\textwidth]{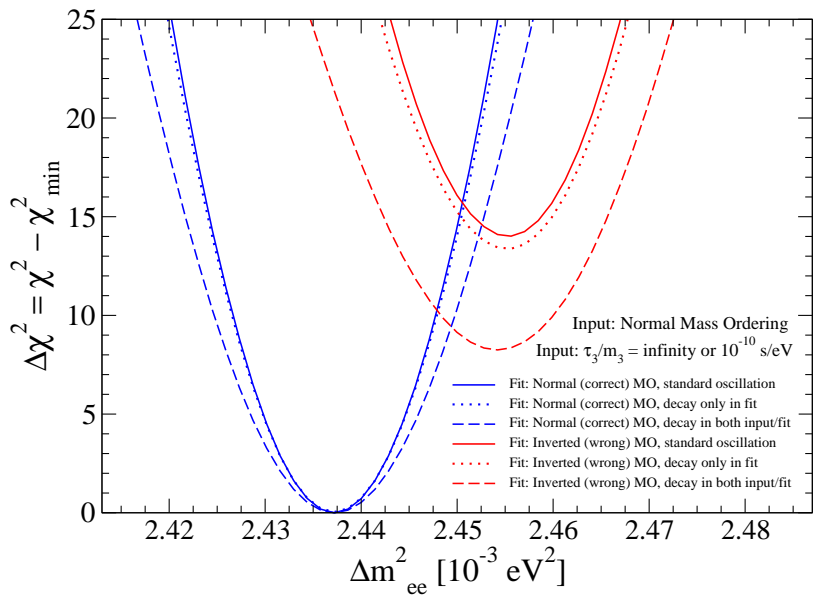}
\end{center}
\vspace{-0.8cm}
\caption{
$\Delta \chi^2 \equiv \chi^2-\chi^2_\text{min}$ 
is shown as a function of the fitted value of $\Delta m^2_\text{ee}$.
The solid blue and red curve correspond, respectively,
to the case where the fit is performed assuming the normal 
(right) and inverted (wrong) mass ordering (MO) for the standard
oscillation. 
The blue (red) dotted and dashed curves corresponds, 
respectively, to the case where the input value of $\tau_3/m_3$ 
is $\infty$ and $10^{-10}$ s/eV for the case of 
normal (inverted) mass ordering. 
}
\label{fig:Delta_chi2-MO}
\end{figure}
%%%%%%%%%%%%%%%% FIG 3 %%%%%%%%%%%%%%%%%%

To know the effects of $\nu_3$ decay on the resolution capability of the mass ordering in JUNO, we compute $\Delta \chi^2 \equiv \chi^2-\chi^2_\text{min}$ by taking both the normal (right) and inverted (wrong) mass ordering.\footnote{
%%%%%%%%%%%%%% footnote %%%%%%%%%%%%%%
At the end of this section~\ref{sec:mass-ordering}, we will place a long clarifying remark on how to interpret our $\Delta \chi^2$. 
}
In Fig.~\ref{fig:Delta_chi2-MO}, $\Delta \chi^2$ is plotted as a
function of $\Delta m^2_\text{ee}$ obtained by marginalizing all the
other parameters. 
For the abscissa in Fig.~\ref{fig:Delta_chi2-MO} we use 
$\Delta m^2_\text{ee} \equiv \vert c^2_{12} \Delta m^2_{31} + s^2_{12}
\Delta m^2_{32} \vert $, which we believe to be the appropriate variable
to discuss resolution of the mass ordering in medium baseline reactor experiments~\cite{Minakata:2007tn}. 
It is proposed as the effective atmospheric $\Delta m^2$ determined by
$\bar{\nu}_e$ disappearance experiment in vacuum~\cite{Nunokawa:2005nx}, 
which agrees in a good approximation with the one measured by the reactor $\theta_{13}$ experiment~\cite{An:2013zwz}.

The blue and the red curves are for the normal (input) and the inverted (wrong) mass orderings, respectively. The three line types, which are common to the both mass orderings, correspond respectively to: 

\noindent
(i) Solid curve: no $\nu$ decay in both the input and the fit, the case of standard oscillation, 

\noindent 
(ii) Dotted curve: no $\nu$ decay in the input but $\nu_3$ decay is allowed with prior-unconstrained lifetime in the fit, and 

\noindent
(iii) Dashed curve: $\nu_3$ decay with lifetime $\tau_3/m_3 =10^{-10}$ s/eV 
is assumed in the input and allowed in the fit. 

The global features of the results can be summarized as:
\begin{itemize}
\item 
From (i) to (ii) there is only minor change (up to $\sim 1$) in $\Delta \chi^2$. It means that allowing the decay only in the fit little affects the output. 

\item 
From (i) to (iii) there are appreciable changes in $\Delta \chi^2$ in a manner which depend very much on the mass ordering. In the case of normal (right) mass ordering the change in $\Delta \chi^2$ upon turning on $\nu_3$ decay both in the input and output is modest, slightly opening up the Gaussian parabola. 

\item 
If $\Delta \chi^2$ difference between the right (input normal) and the
wrong mass orderings without decay is denoted as 
$\Delta \chi^2_{\text{no-decay}}$, 
$\Delta \chi^2$ difference at the minima across the different 
mass orderings becomes $\simeq \Delta \chi^2_{ \text{no-decay} } -5$
when the $\nu_3$ decay is turned on with lifetime 
comparable to the current and the JUNO bound, $\tau_3/m_3 =10^{-10}$ s/eV. 

\end{itemize}

\noindent
Thus, we have observed that the $\nu_3$ decay has a big impact on mass ordering resolution in JUNO, significantly worsening the sensitivity. 

The readers must be noticed that we {\em carefully avoided } to make a
quantitative statement on how large is $\Delta \chi^2_{\text{no-decay}}$,
but restricted ourselves into the change due to the decay effect. 
It is likely that our procedure to simulate the distortion of the event number energy distribution is insufficient to reliably extract the absolute confidence level for the mass ordering determination. In particular, we do not take into account the uncertainties related to the non-linearity of the energy measurement, whose control would be the key to the success in the mass ordering measurement in JUNO.\footnote{
%%%%%%%%%%%%% footnote %%%%%%%%%%%%%%%
This issue was raised in \cite{Minakata:2007tn}, and the crucial importance of controlling the energy-scale errors on mass ordering determination was illuminated with experimental perspectives by the authors of ref.~\cite{Qian:2012xh}. 
}
On the other hand, the omitted error in computing $\chi^2$ may not affect our discussion of the effect of decay in a significant way, because it affects the spectrum as simple reduction of the oscillation amplitude.\footnote{
%%%%%%%%%%%%% footnote %%%%%%%%%%%%%%%
In addition, there is a subtle issues related to statistical treatment
of mass ordering determination. 
In the usual case one can attribute to $\sqrt{\Delta \chi^2_\text{min}}$ the
meaning of $1 \sigma$ significance. The special feature that the fitting 
parameter is a discrete variable requires a different treatment to
evaluate correctly the CL for the mass ordering 
determination~\cite{Qian:2012zn,Ciuffoli:2013rza,Capozzi:2013psa}. 
Since we do not intend to elaborate this point, and because of 
the crude nature of our $\Delta \chi^2$ construction, we recommend 
the readers to use the information presented in 
Fig.~\ref{fig:Delta_chi2-MO} only to have a feeling on 
how $\nu_3$ decay affects the determination of the mass ordering in JUNO. 
}

\subsection{Impact of decay on the determination of the oscillation parameters}
\label{subsec:decay-impact-mass-mixing}

We briefly discuss in this section possible effects of $\nu_3$ decay on determination of the standard oscillation parameters in JUNO as well as on the values of output uncertainties. Detailed features of parameter correlations with and without $\nu_3$ decay, including the correlations with the systematic errors, will be discussed in appendix~\ref{sec:correlation}. As in the previous section~\ref{sec:mass-ordering}, we consider the three cases, (i) standard oscillation, (ii) decay effect only in the fit and (iii) decay effect both in the input and in the fit. For the finite input value of lifetime we always use $\tau_3/m_3 = 10^{-10}$ s/eV. 

\begin{figure}[H]
\begin{center}
\vspace{-1.6cm}
\includegraphics[bb=0 0 842 595,width=0.95\textwidth]{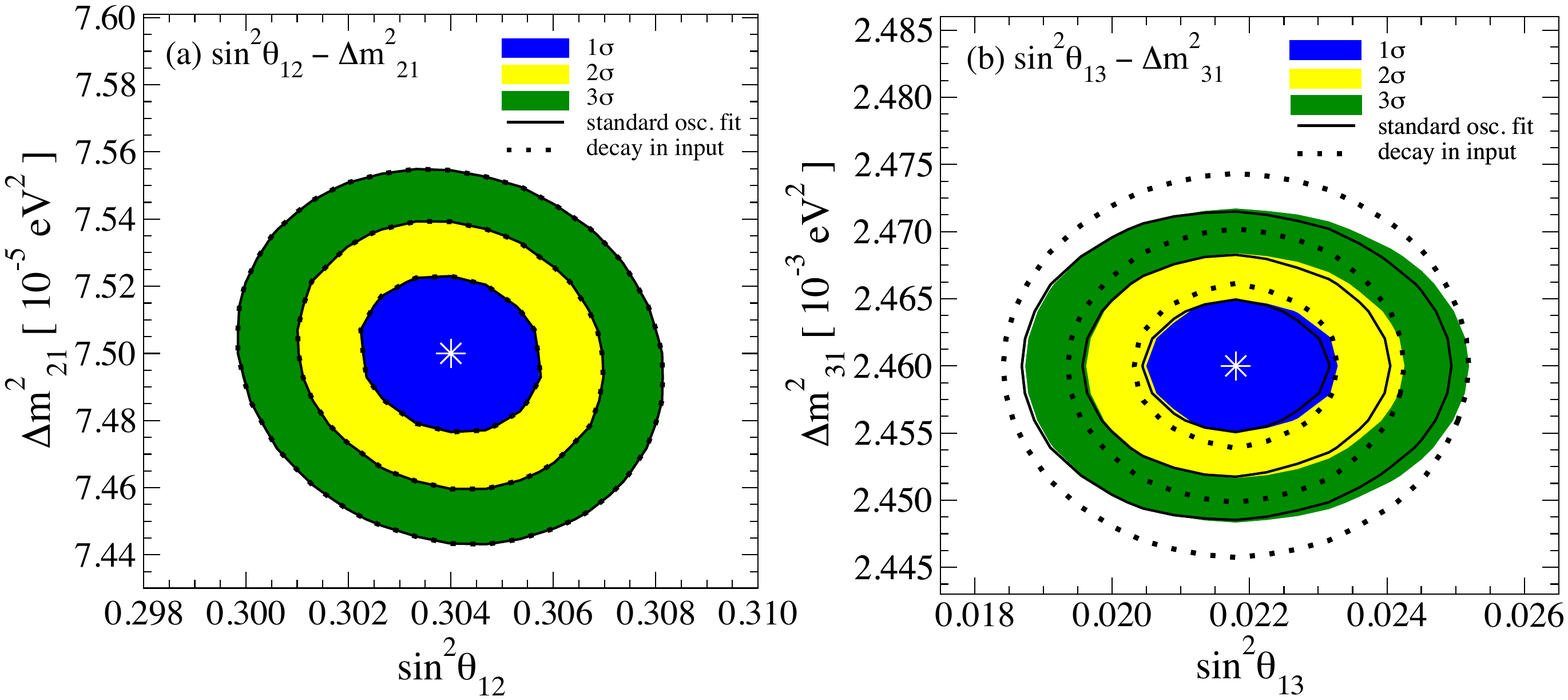}
\end{center}
\vspace{-2.8cm}
\caption{
The regions allowed at 1, 2, and 3$\sigma$ CL are drawn, respectively, 
by the filled colors of blue, yellow and green 
in the parameter space spanned by  
(a) $\sin^2 \theta_{12}$ - $\Delta m^2_{21}$ (left panel)
and (b) $\sin^2 \theta_{13}$ - $\Delta m^2_{31}$ (right panel)
for the case (ii) where no decay is assumed in input but allowed in the fit. 
For comparison, the cases of (i) the standard oscillation fit without decay
and (iii) the case with the input value of $\tau_3/m_3 = 10^{-10}$ s/eV 
are shown, respectively, by the black solid and black dotted curves. 
5 years of exposure is assumed. }
%}
\label{fig:allowed-standard-param}
\end{figure}

To present our results in a clear way, we always use the following line
symbols throughout figures~\ref{fig:allowed-standard-param} (this
section), \ref{fig:correlation-decay} (appendix~\ref{sec:decay-13}), and
\ref{fig:allowed-regions-others}
(appendix~\ref{sec:miscellaneous}):   
For the case (ii) we show the 1, 2 and 3$\sigma$ allowed regions, respectively, by the filled blue, yellow and green colors. For the case (i) the allowed regions are delimited by the black solid curves, and for (iii) by the black dotted curves. 

In Fig.~\ref{fig:allowed-standard-param} we show the allowed regions of the parameters in (a) $\sin^2\theta_{12}$ - $\Delta m^2_{21}$ (left panel), and (b) $\sin^2\theta_{13}$ - $\Delta m^2_{31}$ (right panel) space. 
We observe in the left panel that there is no visible effect of the decay on the determination of the parameters in the 1-2 sector of the MNS matrix, $\sin^2\theta_{12}$ and $\Delta m^2_{21}$. Whereas in the right panel, it is revealed that accuracy in measurement of $\Delta m^2_{31}$ (of $\sin^2 \theta_{13}$ also but only slightly) is affected when $\nu_3$ decays. 
%%%%%%%% I AVOIDED FURTHER COMMENTS WHICH ARE LEFT FOR ITEMIZED ONES
%
%%%%%%%%%%%%%%%% FIG 4 %%%%%%%%%%%%%%%%%%
%%%%%%%%%%%%%%%% FIG 4 %%%%%%%%%%%%%%%%%%

%%%%%%%%%%%%%%%%%%%%%%%%%%%%%%%%%%%%%%%
\begin{table}[h!]
\vglue -0.2cm
\begin{center}
\caption{
Fractional errors in percent (at $1\sigma$) 
of the oscillation as well as the parameters for systematic uncertainties, 
before (prior) and after the fit to 5 years of JUNO data 
with 100\% detection efficiency, 
for the case where the true mass ordering is normal. 
They are determined by the condition $\Delta \chi^2 = 1$ 
for one degree of freedom. 
We consider three cases 
(i) standard oscillation fit without decay
(ii) decay sensitivity fit (with true $\tau_3/m_3 =\infty$)
and (iii) decay measurement fit (with true $\tau_3/m_3 =10^{-10}$ s/eV).
}
\label{tab:frac-uncert} 
\vglue 0.2cm
\begin{tabular}{c|c|ccc}
\hline 
parameter & prior error (\%)
 &  & fitted error (\%) & \\
&  & (i) & (ii)  & (iii)  \\
\hline 
$\sin^2\theta_{12}$ & 4.1 & 0.35& 0.35 & 0.35\\
$\Delta m^2_{12}$ & 4.1 & 0.21 & 0.21& 0.21\\
$\sin^2\theta_{13}$ & 4.6 & 3.7 &3.8 & 4.3 \\
$\Delta m^2_{13}$ & 1.9 & 0.12 & 0.12& 0.16 \\
\hline 
$1+\xi_\text{reac}$ & 3.0 & 0.50 & 0.50& 0.51\\
$1+\xi_\text{U}$ & 20 & 12 & 12& 12\\
$1+\xi_\text{Th}$ & 20 & 13 & 13& 13\\
$1+\eta$ & 10 & 5.5 & 6.0 & 7.1\\
\hline 
\end{tabular}
\end{center}
\vglue -0.4cm
\end{table}
%%%%%%%%%%%%%%%%%%%%%%%%%%%%%%%%%%%%%%%

To obtain a more comprehensive view of effect of decay and to represent the change in the sensitivities in a more quantitative way, we present in Table \ref{tab:frac-uncert} the fractional uncertainties (in \%) of all the parameters determinable in the three cases (i-iii) in the JUNO setting. In doing so we do not restrict to the 
oscillation parameters, but also include the ones which describe the systematic errors. From Table \ref{tab:frac-uncert} and Fig.\ref{fig:allowed-standard-param} we observe:
\begin{itemize}

\item 
Comparison between the columns (i) and (ii) indicates that if the decay effect is considered only in the fit, its impact is virtually absent except for $\theta_{13}$ and the energy resolution uncertainty parameter $\eta$. 

\item 
Comparison between the columns (i) and (iii) tells us that if the decay effect is considered for both in the input and the fit, there is appreciable impact of the size of approximately 10\%-30\% but only for $\sin^2\theta_{13}$, $\Delta m^2_{31}$ and 1+$\eta$. 

\end{itemize}
\noindent
As a whole, we conclude that the impact of the decay 
on the determination of the mass and mixing parameters 
is rather small.

\section{Conclusions}
\label{sec:conclusions}

In this paper, we have investigated the question of how strong 
a constraint on  decay lifetime of the massive neutrino state $\nu_{3}$ 
can be placed by the medium-baseline ($L \sim 50$ km) 
reactor neutrino experiments, JUNO or RENO-50, 
which we referred simply as JUNO in most part in the text. 
Assuming decay into invisible states and $\tau_{1}, \tau_{2} \gg
\tau_{3}$, we found that the bound 
$\tau_3/m_3 > 7.5 \ (5.5)\ \times 10^{-11}$ s/eV at 95\% (99\%) CL
can be obtained by JUNO with its five years exposure with 100\% efficiency.

In fact, there is a simple reason why the JUNO setting can offer the chance of deriving the most stringent bound on the neutrino lifetime among all the experiments which utilize neutrinos from the artificial sources. 
Given that such experiments are designed to have sensitivities at around the first oscillation maximum (in exploring the atmospheric-scale neutrino oscillations), the kinematic estimate implies ${ \tau }/{ m } \sim { L }/{ E } \simeq { 2 \pi }/{ \Delta m^2_\text{atm} }$. However, if there is such an experiment that can explore atmospheric-scale oscillations at the distance of solar-scale oscillation maximum, the bound on ${ \tau }/{ m }$ would become severer by a factor of ${ \Delta m^2_\text{atm} }/{ \Delta m^2_{21} } \simeq 30$. This is what JUNO does. 

The bound on ${ \tau_3 }/{ m_3 }$ we obtained for 5 years exposure of JUNO has the same order of magnitude as (but is somewhat weaker than) the one obtained with the atmospheric neutrino data. We have examined the question of whether the 15 years running of JUNO can tighten up the bound to the level of the current atmospheric neutrino bound, and obtained an affirmative answer. It is nice to see that the comparably strong $\nu_3$ lifetime bound can be deduced by the two completely different experiments, one observing reactor neutrinos at the medium baseline, and the other measuring the atmospheric neutrinos.

We have also discussed to what extent the $\nu_{3}$ decay could affect the determination of the mass ordering as well as the precision parameter measurement by JUNO. We considered the two cases (in the notation in section~\ref{sec:impact-decay}), (ii) the decay effect is absent in the input data set but allowed in the output (fit), and (iii) $\nu_3$ decay in input with the lifetime $\tau_3/m_3 = 10^{-10}$ s/eV, which is marginally consistent with the current and our JUNO bound. For the mass ordering determination, we found the significant impact of the decay (reduction of $\Delta \chi^2$ by $\sim$ five units) but only for the case (iii). With regard to influence of decay to the measurement of the oscillation parameters, we found that the impact is rather small except for $\Delta m^2_{31}$ again only for the case (iii), in which the uncertainty of $\Delta m^2_{31}$ would become $\sim$ 30\% larger.

\appendix

\section{Neutrino oscillation probabilities in the presence of neutrino decay in vacuum}
\label{sec:probabilities}

The neutrino evolution in vacuum with neutrino decay is governed by the usual Schr\"odinger equation by replacing the neutrino energy $E_i$ by the one in (\ref{ith-energy}). The $S$ matrix whose elements describe neutrino flavor transition in vacuum as $\nu_{\alpha} (x) = S_{\alpha \beta} (x) \nu_{\beta} (0)$ is given by 
\begin{eqnarray}
S_{\alpha \beta} (x) = \sum_{i=1}^{3}
U_{\alpha i} U_{\beta i}^* 
e^{ - i \frac{m_{i}^2 }{2E} x } e^{ - \frac{ \Gamma_{i} }{2} x },
\label{amplitude}
\end{eqnarray}
where $x$ is the distance traveled by neutrino and the notations for $U_{\alpha i}$, $m_i$, $E$, and $\Gamma_{i}$ are the same as in section~\ref{sec:uniqueness}.
Then, the general expression of the three-flavor expression of 
oscillation probability with neutrino decay in vacuum is given by \cite{Lindner:2001fx}
\begin{eqnarray}
\hskip -1.5cm 
& & P(\nu_\beta \rightarrow \nu_\alpha) \equiv \vert S_{\alpha \beta}\vert
= P(\nu_\beta \rightarrow \nu_\alpha: \text{no decay} )
- \sum_{i} \vert U_{\alpha i} \vert^2 \vert U_{\beta i} \vert^2 
\left( 1 - e^{- \Gamma_{i} x } \right) 
\nonumber \\ 
\hskip -1.5cm 
&-& 
2 \sum_{j > i}  
\text{Re}[{\cal C}_{\alpha \beta i j}]
\cos\Delta_{ji}
\left( 1 - e^{ - \frac{ \Gamma_{i} + \Gamma_{j}  }{2} x } \right) 
+
2 \sum_{j > i} 
\text{Im}[{\cal C}_{\alpha \beta i j}]
\sin \Delta_{ji}
\left( 1 - e^{ - \frac{ \Gamma_{i} + \Gamma_{j}  }{2} x } \right), 
\label{general-probability}
\end{eqnarray}
where 
${\cal C}_{\alpha \beta i j} \equiv 
U_{\alpha i} U_{\beta i}^* U_{\alpha j}^* U_{\beta j}$,
$\Delta_{ji} \equiv \Delta m_{ji}^2 x/(2 E)$. 
The first term in (\ref{general-probability}), the oscillation probability in the absence of neutrino decay, 
is given by the familiar expression,
\begin{eqnarray}
\hskip -1cm 
P(\nu_\beta \rightarrow \nu_\alpha: \text{no decay} ) 
= \delta_{\alpha \beta} 
- 4 \sum_{j > i}  
\text{Re}[{\cal C}_{\alpha \beta i j}] 
\sin ^2 \left( \frac{\Delta_{ji}}{2}\right)
- 2 \sum_{j > i} 
\text{Im}[{\cal C}_{\alpha \beta i j}] 
\sin \Delta_{ji}.
\label{P-beta-alpha-no-decay}
\end{eqnarray}

If the approximations 
$\Gamma_{1} x \sim \Gamma_{2} x \ll1$, and $\Gamma_{3} x \sim  1$ 
holds we can simply (\ref{general-probability}). 
The disappearance oscillation probability takes the particularly simple form as
\begin{eqnarray}
&& P(\nu_{\alpha} \rightarrow \nu_{\alpha} ) = 
1 - 4 \vert U_{\alpha 1} \vert^2 \vert U_{\alpha 2} \vert^2
\sin^2 \left( \frac{\Delta_{21}}{2} \right)
- 4 \sum_{i=1,2}  \vert U_{\alpha i} \vert^2 \vert U_{\alpha 3} \vert^2
\sin^2 \left( \frac{\Delta_{3i}}{2} \right)
\nonumber \\ 
&-& 
\vert U_{\alpha 3} \vert^4 
\left( 1 - e^{- \Gamma_{3} x } \right)  
- 2 \sum_{i = 1, 2} 
\vert U_{\alpha i} \vert^2 \vert U_{\alpha 3} \vert^2 
\cos \Delta_{3i} 
\left( 1 - e^{ - \frac{ \Gamma_{3}  }{2} x } \right),  
\nonumber \\ 
&\approx& 
1 - 4 \vert U_{\alpha 1} \vert^2 \vert U_{\alpha 2} \vert^2
\sin^2 \left( \frac{\Delta_{21}}{2} \right)
- 4 \vert U_{\alpha 3} \vert^2 
\left( 1 -  \vert U_{\alpha 3} \vert^2 \right) 
\sin^2 \left( \frac{\Delta_\text{atm}}{2} \right)
\nonumber \\ 
&-& 
\vert U_{\alpha 3} \vert^4 
\left( 1 - e^{- \Gamma_{3} x } \right)  
- 2 \vert U_{\alpha 3} \vert^2 
\left( 1 -  \vert U_{\alpha 3} \vert^2 \right) 
\cos \Delta_\text{atm}
\left( 1 - e^{ - \frac{ \Gamma_{3}  }{2} x } \right),  
\label{P-alpha-alpha}
\end{eqnarray}
where in the last line we have used the approximation 
$\Delta_{31} \approx \Delta_{32} \equiv \Delta_\text{atm}$.\footnote{
%%%%%%%%%%%%% footnote %%%%%%%%%%%%%%%%%
If we were to go through a theoretical discussion of the effect of neutrino decay on determination of the neutrino mass ordering, we should have made our treatment more elaborate one. It would necessitate use of the effective atmospheric $\Delta m^2_\text{ee}$ as mentioned earlier. However, determination of mass ordering is little affected by the decay effect as far as $\tau_3/m_3$ is much smaller than the current bound. Though there is some influence if the true value of $\tau_3/m_3$ is comparable with the current bound, we prefer to remain in the simpler treatment above in this paper. 
If necessary, it is straightforward to formulate this problem 
by taking the framework adopted in \cite{Minakata:2007tn}. 
%%%%%%%%%%%%% footnote %%%%%%%%%%%%%%%%%
}
In the appearance channels ($\alpha \neq \beta$), the oscillation probability also simplifies:  
\begin{eqnarray}
&&P(\nu_\beta \rightarrow \nu_\alpha) 
=
P(\nu_\beta \rightarrow \nu_\alpha: \text{no decay} )
- \vert U_{\alpha 3} \vert^2 \vert U_{\beta 3} \vert^2 
\left( 1 - e^{- \Gamma_{3} x } \right) 
\nonumber \\ 
&-& 
2 \sum_{i = 1, 2} 
\text{Re}[{\cal C}_{\alpha \beta i j}] 
\cos \Delta_{3i}
\left( 1 - e^{ - \frac{ \Gamma_{3} }{2} x } \right) 
%
%\nonumber\\
%&+&
+
2 \sum_{i = 1, 2} 
\text{Im}[{\cal C}_{\alpha \beta i j}] 
\sin \Delta_{3i}
\left( 1 - e^{ - \frac{ \Gamma_{3} }{2} x } \right) 
\nonumber \\ 
&\approx& 
P(\nu_\beta \rightarrow \nu_\alpha: \text{no decay} )
- \vert U_{\alpha 3} \vert^2 \vert U_{\beta 3} \vert^2 
\left( 1 - e^{- \Gamma_{3} x } \right) 
\nonumber \\ 
&-& 
2 \vert U_{\alpha 3} \vert^2 \vert U_{\beta 3} \vert^2  
\cos \Delta_\text{atm} 
\left( 1 - e^{- \frac{ \Gamma_{3} x}{2} } \right)
\label{P-beta-alpha}
\end{eqnarray}
Under the approximation $\Delta_{31} \approx \Delta_{32} \equiv
\Delta_\text{atm}$
and because of 
anti-symmetry of 
$\text{Im}[U_{\alpha i} U_{\beta i}^* U_{\alpha j}^*U_{\beta j}]$ 
$(=\text{Im}[{\cal C}_{\alpha \beta i j}])$ 
under $i \leftrightarrow j$, 
CP violating term in the decay-width dependent part of the oscillation probabilities vanishes. 

Some remarks are in order:

\begin{itemize}

\item 
In $\nu_{\mu}$ disappearance channel the coefficients of 
$\left( 1 - e^{- \Gamma_{3} x } \right)$ and 
$\cos \Delta_\text{atm}$ terms 
are given by $\vert U_{\mu 3} \vert^4 = s^4_{23} c^4_{13} \simeq 0.24 $ and $2 \vert U_{\mu 3} \vert^2 \left( 1 -  \vert U_{\mu 3} \vert^2 \right) \simeq 0.50$, respectively. Therefore, the oscillation independent $\nu_{\mu}$ depletion effect may be as important as the effect of diminishing amplitude in the atmospheric-scale oscillations.

\item 
In $\nu_{e}$ appearance channel the corresponding coefficients of the
      two 
$\Gamma_{3}$-dependent terms are given by $\vert U_{e 3} \vert^2 \vert U_{\mu 3} \vert^2 = s^2_{23} c^2_{13} s^2_{13} \simeq 1.1 \times 10^{-2} $ and twice of that $\simeq 2.2 \times 10^{-2}$, respectively. They are comparable with each other, and are similar in magnitude as the main oscillation term in vacuum. 

\end{itemize}

\section{Event energy spectrum}
\label{sec:spectrum}

The distribution of the number of events 
coming from the inverse $\beta$-decay (IBD) reaction, 
$\bar{\nu}_e + p \to e^+ + n$,
as a function of the visible energy is given by, 
\begin{eqnarray}
\frac{dN(E_{\text{vis}})}{dE_\text{vis}} 
&= & n_p t_\text{exp} \int_{m_e}^\infty E_e
\int_{E_\text{min}}^\infty dE 
\sum_{i = \text{reac}, \text{geo}-\nu}
\frac{d\phi_i(E)}{dE}
\epsilon_{\text{det}}(E_e)
\frac{d\sigma(E_\nu,E_e)}{dE_e} \nonumber \\
& & \times 
P_i(\bar{\nu}_e \to \bar{\nu}_e; L_i, E) 
R(E_e,E_{\text{vis}}),
\label{eq:event-dist}
\end{eqnarray}
where $n_p$ is the number of target (free protons), 
$t_\text{exp}$ is the exposure, 
$\epsilon_{\text{det}}$ is the detection 
efficiency, 
$d\phi_i(E)/dE$ is the differential fluxes 
of reactor neutrinos and geoneutrions, 
$d\sigma(E_\nu,E_e)/dE_e$ is the IBD cross section, 
$P_i(\bar{\nu}_e \to \bar{\nu}_e; L_i, E)$
is the $\bar{\nu}_e$ survival probabilities
and $R(E_e,E_{\text{vis}})$ is the Gaussian resolution 
function (see below).  

For simplicity, we set $\epsilon_{\text{det}} =1$, 
and as a reasonable approximation 
for our purpose, we ignore the neutron recoil 
in the IBD reaction and simply assume that
neutrino energy, $E$, and the positron energy, $E_{e}$, 
is related as $E_{e} = E - (m_n-m_p) \simeq E - 1.3$ MeV. 
Due to the finite energy resolution, 
the event distribution can not be obtained 
as a function of $E_{e}$ (true positron energy) 
but as a function of the reconstructed or so called 
visible energy, $E_\text{vis}$, which is approximately related 
to neutrino energy as 
$E_\text{vis} \simeq E - (m_n-m_p) + m_e$,
after taking into account the energy resolution
(see the text below). 
Regarding the cross section, $d\sigma(E_\nu,E_e)/dE_e$, 
we use the one found in \cite{Strumia:2003zx}. 

For the JUNO detector, we assume 
the same proton fraction $\simeq 11$\% as 
the Daya Bay detectors ~\cite{DayaBay:2012aa}
which implies $\sim 1.44 \times 10^{33}$ 
free protons for 20 kt. 
The differential flux of reactor neutrino 
${d\phi(E)}/{dE}$ can be computed as, 
\begin{eqnarray}
\frac{d\phi(E)}{dE}   
= \frac{1}{4 \pi L^2}   
S(E) \frac{P_\text{th}}{ \langle E \rangle},
\end{eqnarray}
where $P_\text{th}$ is the thermal power of the reactor,  
${\langle E \rangle} \simeq $ 210 MeV is the average
energy released by per fission 
computed by taking into account the ratios
of the fuel compositions of the reactor (see below). 

We can replace, in a good approximation, 
the reactor complex consisting of 6 and 4 reactors, respectively, 
at Yangjiang and Taishan sites by a single reactor with 
the thermal power of 35.8 GW placed at the baseline 
$L=52.5$ km from the JUNO detector. 
We also include the contributions from 
the far reactor complexes at Daya Bay (with the baseline of 215 km)
and Huizhou (with the baseline of 265 km) sites, 
which contribute, respectively, about 3\% and 2\% 
in terms of the total number of events. 

For the reactor spectra $S(E)$ we use the convenient
analytic expressions found in ~\cite{Mueller:2011nm}
with the typical fuel compositions of the reactors, 
$^{235}$U: $^{239}$Pu: $^{238}$U: $^{241}$Pu 
= 0.59: 0.28: 0.07: 0.06, obtained by 
taking time average of Fig. 21 of \cite{An:2013uza}. 
We note that $S(E)$ is nothing but the 
number of neutrinos being emitted per fission per energy (MeV). 

Furthermore, we also include, for completeness, 
geoneutrinos coming from the decays of U and Th inside the earth in 
the same way as done in \cite{Capozzi:2013psa}, 
despite that it is not important for our main purpose. 
Assuming the input (true) geoneutrinos fluxes 
of 
$\phi(\text{U}) = 4.0\times 10^6$ cm$^{-2}$ s$^{-1}$
and 
$\phi(\text{Th}) = 3.7\times 10^6$ cm$^{-2}$ s$^{-1}$, 
the expected U and Th geoneutrinos induced events 
are, respectively, $\sim 1.9 \times 10^3$ and 
$\sim 5.4 \times 10^2$ for 5 years of exposure at JUNO
detector. 
For simplicity,
we consider only the averaged standard oscillation 
for geoneutrinos, and ignore the decay effect for them.
Since the presence of geoneutrinos 
has minor impact on the standard oscillation study at JUNO, 
their decay effect is even minor and we believe
that this is a fairly good approximation. 
In fact we have verified that the presence of geoneutrinos 
has virtually no impact on the sensitivity to the decay effect. 
In the fit, we let the fluxes of geoneutrinos vary 
freely subject to the pull terms in Eq. (\ref{eq:chi2_sys}) 
with $\sigma_{\xi_\text{U}} = \sigma_{\xi_\text{Th}} = $ 20\%.

$R(E_e,E_{\text{vis}})$ is the function 
which takes into account the finite energy resolution 
of the detector and is given by 
\begin{eqnarray}
R(E_e,E_{\text{vis}})
\equiv 
\frac{1}{\sqrt{2\pi}\sigma(E_e)}
\text{exp}
\left[ -\frac{1}{2}
\left( \frac{E_e+m_e-E_{\text{vis}}}
{\sigma(E_e)}\right)^2
\right]
\label{eq:resolution-function}
\end{eqnarray}
where the energy resolution is assumed to be~\cite{Li:2013zyd}, 
\begin{eqnarray}
\frac{\sigma(E_e)}{(E_e+m_e)} = 
\frac{3\%\ (1+\eta)}{\sqrt{(E_e+m_e)/\text{MeV}}}, 
\label{Eresolution}
\end{eqnarray}
where $\eta$ is introduced to take into account the uncertainty in the energy 
resolution. The energy resolution in (\ref{Eresolution}) amount to consider only the stochastic term.

The expected total number of events at JUNO for the 5 years of exposure 
with 100\% detection efficiency is $1.41 \times 10^5$. 
The individual contributions from medium-baseline reactor sites 
(at Yangjiang and Taishan), far reactor sites (at Daya Bay and Huizhou), 
geoneutrinos are, respectively, 1.39$\times 10^5$, 
6.70$\times 10^3$ and 2.40$\times 10^3$.

\section{Correlations among the oscillation, systematic and decay parameters}
\label{sec:correlation}

Here, we discuss the correlations among the mass and the mixing parameters, systematic uncertainty parameters, and the decay parameter $\tau_3/m_3$. For this purpose we have examined all possible correlations among the mixing parameters and the uncertainty parameters.\footnote{
%%%%%%%%%%%%%% footnote %%%%%%%%%%%%%%%%
They are the following 32 combinations 
(to be shown in Fig.~\ref{fig:allowed-regions-others})
in addition to those already shown in
Fig.\ref{fig:allowed-standard-param} 
and the one to be shown in Fig.~\ref{fig:correlation-decay}:\\ 
(a) $\sin^2\theta_{12}$-$\sin^2\theta_{13}$, (b) $\sin^2\theta_{12}$-$\Delta m^2_{31}$, 
(c) $\sin^2\theta_{13}$-$\Delta m^2_{21}$, (d) $\Delta m^2_{31}$-$\Delta m^2_{21}$, \\
(e) $\sin^2\theta_{12}$-$(1+\xi_\text{reac})$,  (f) $\sin^2\theta_{13}$-$(1+\xi_\text{reac})$, 
(g) $\Delta m^2_{21}$-$(1+\xi_\text{reac})$, (h) $\Delta m^2_{31}$-$(1+\xi_\text{reac})$, \\
(i) $\sin^2\theta_{12}$-$(1+\xi_\text{U})$, (j) $\sin^2\theta_{13}$-$(1+\xi_\text{U})$, 
(k) $\Delta m^2_{21}$-$(1+\xi_\text{U})$, (l) $\Delta m^2_{31}$-$(1+\xi_\text{U})$, \\
(m) $\sin^2\theta_{12}$-$(1+\xi_\text{Th})$, (n) $\sin^2\theta_{13}$-$(1+\xi_\text{Th})$, 
(o) $\Delta m^2_{21}$-$(1+\xi_\text{Th})$, (p) $\Delta m^2_{31}$-$(1+\xi_\text{Th})$, \\
(q) $\sin^2\theta_{12}$-$\sigma_E(1+\eta)$, (r) $\sin^2\theta_{13}$-$\sigma_E(1+\eta)$, 
(s) $\Delta m^2_{21}$-$\sigma_E(1+\eta)$, (t) $\Delta m^2_{31}$-$\sigma_E(1+\eta)$, \\
(u) $(1+\xi_\text{U})$-$(1+\xi_\text{reac})$, (v) $(1+\xi_\text{Th})$-$(1+\xi_\text{reac})$
(w) $(1+\xi_\text{U})$-$(1+\xi_\text{Th})$, (x) $\sigma_E(1+\eta)$-$(1+\xi_\text{reac})$\\
(y) $(1+\xi_\text{U})$-$\sigma_E(1+\eta)$, (z) $(1+\xi_\text{Th})$-$\sigma_E(1+\eta)$
(A) $\sin^2\theta_{12}$-$\tau_3/m_3$, (B) $\Delta m^2_{21}$-$\tau_3/m_3$, \\
(C) $\Delta m^2_{31}$-$\tau_3/m_3$, (D) $(1+\xi_\text{reac})$-$\tau_3/m_3$, 
(E) $(1+\xi_\text{U})$-$\tau_3/m_3$ and (F) $(1+\xi_\text{Th})$-$\tau_3/m_3$.
}
The calculation is done under the same conditions (running time etc.) as assumed for calculating the lifetime bound given in Fig.~\ref{fig:Delta_chi2}. 

The general features of the correlations among the oscillation parameters, systematic uncertainty parameters and the decay parameter ($m_3/\tau_3$) can be summarized as follows:
\begin{itemize}
\item 
The effect of $\nu_3$ decay is visible only in 
the contours which involve the energy resolution 
$\sigma_{E} (1+\eta)$, $\theta_{13}$
and $\Delta m^2_{31}$, 
as implied by the results shown in Table~\ref{tab:frac-uncert}.\footnote{
%%%%%%%%%%%%% footnote %%%%%%%%%%%%%%%
If we remove all the priors for the mixing parameters (i.e., pull terms) 
the feature changes dramatically. There arises an extended allowed
region up to $\sin^2 \theta_{13} \simeq 0.044$ at 2$\sigma$ C.L. with 
a larger flux normalization by about 5\%.}

\item 
The correlations between $\tau_3/m_3$ and 
$\sigma_{E} (1+\eta)$, $\theta_{13}$ exist, but in easily understandable way.

\end{itemize}

\subsection{Decay lifetime vs. $\theta_{13}$ and the energy resolution}
\label{sec:decay-13}

In this subsection, we highlight only the correlations between the $\nu_3$ lifetime and (a) $\sin^2 \theta_{13}$ and (b) the uncertainty in energy resolution. All the other combinations not shown here and in section~\ref{subsec:decay-impact-mass-mixing} will be shown in the next subsection, see Fig.~\ref{fig:allowed-regions-others}. 

%%%%%%%%%%%%%%%% FIG 5 %%%%%%%%%%%%%%%%%%
\begin{figure}[h!]
\begin{center}
\vspace{-1.4cm}
\includegraphics[bb=0 0 842 595,width=0.95\textwidth]{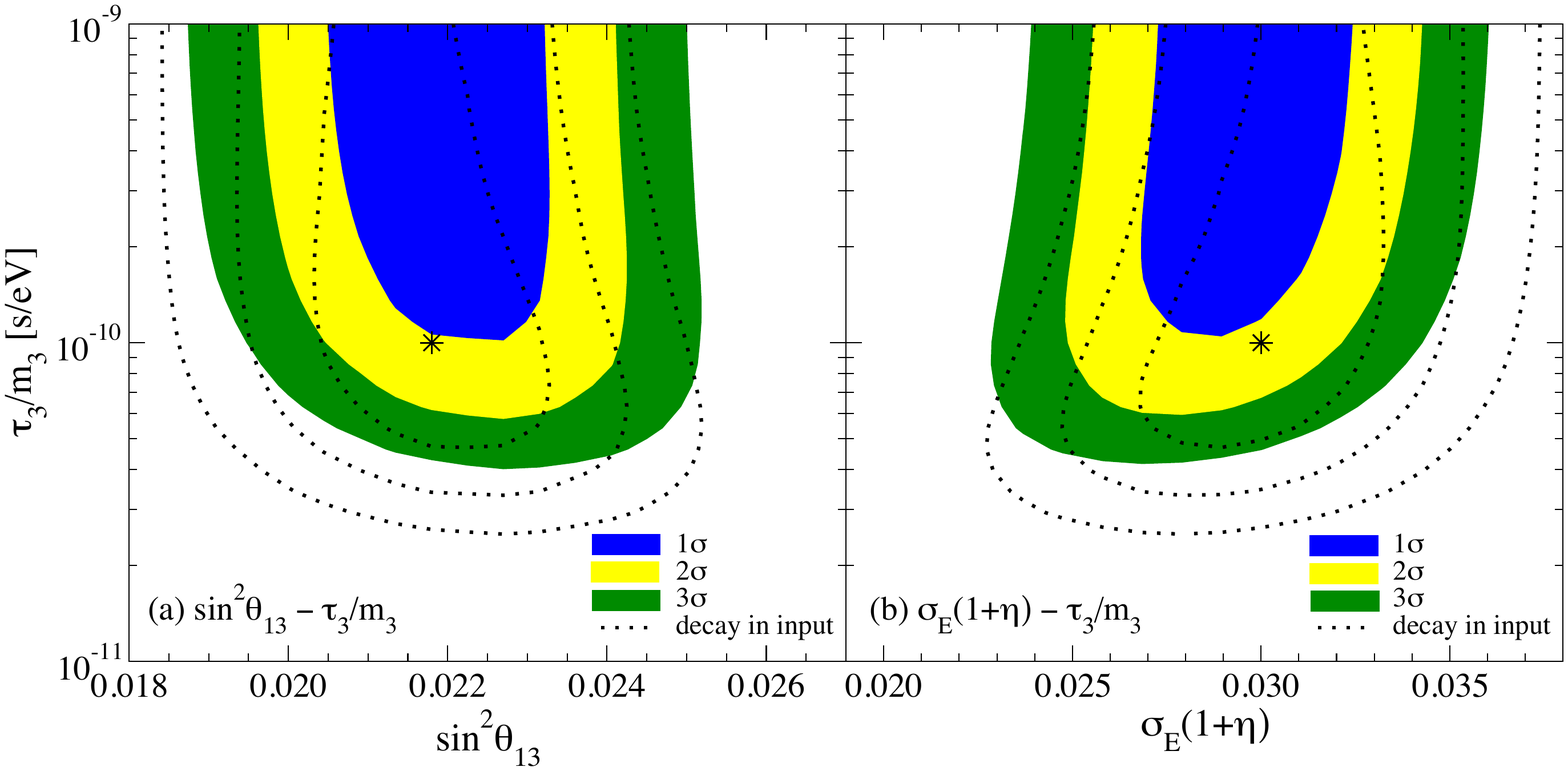}
\end{center}
\vspace{-2.5cm}
\caption{
The regions allowed at 1, 2, and 3$\sigma$ CL are drawn, respectively, 
by the filled colors of blue, yellow and green 
in the parameter space spanned by  
(a) $\sin^2 \theta_{13}$ - $\tau_3/m_3$ (left panel)
and (b) $\sigma_E(1+\eta)$ - $\tau_3/m_3$ (right panel) for the 
case where no decay is assumed in input. 
The case where $\tau_3/m_3 = 10^{-10}$ s/eV is assumed 
in input is also shown by the black dotted curves. 
5 years of exposure is assumed. }
\label{fig:correlation-decay}
\end{figure}
%%%%%%%%%%%%%%%% FIG 5 %%%%%%%%%%%%%%%%%%

In Fig.~\ref{fig:correlation-decay}, we show the correlations in space spanned by (a) $\sin^2 \theta_{13}$ - $\tau_3/m_3$ (left panel) and (b) $\sigma_{E}(1+\eta)$ - $\tau_3/m_3$ (right panel). We notice that there are correlations, i.e., the effect of decay, both in the left and right panels, but only in near the lower end of the allowed region of the lifetime. 
From the left panel of Fig.~\ref{fig:correlation-decay} 
we learn that the decay effect (diminishing the amplitude of the
wiggles) can be compensated to some extent by making $\theta_{13}$
larger, which produces a new allowed region toward 
the large $\theta_{13}$ direction. 

Corresponding to the newly emerged region,
there also arises a new allowed region in 
the $\sigma_{E}(1+\eta)$ - $\tau_3/m_3$ space,
as seen in the right panel of Fig.~\ref{fig:correlation-decay}.
It is located at near the top of the peninsula, 
a slightly distorted region toward small $\sigma_{E}(1+\eta)$ direction
in the right panel of Fig.~\ref{fig:correlation-decay}. 
The shape reflect a better energy resolution in the newly allowed region.
We note that due to the above $\sin^2 \theta_{13}$ -
$\sigma_{E}(1+\eta)$ correlation, 
there always exist a slightly extended allowed region in 
any one of the correlation plots which involve 
$\sin^2 \theta_{13}$ or $\sigma_{E}(1+\eta)$ toward 
the direction of larger $\theta_{13}$ and smaller $\sigma_{E}(1+\eta)$.

\subsection{ Correlations of parameters: miscellaneous cases} 
\label{sec:miscellaneous}

We now show, for completeness, the allowed regions spanned by all 
the remaining 32 combinations of parameters considered 
in this work (which are described explicitly in the footnote 12), 
except for those already shown in 
Figs.~\ref{fig:allowed-standard-param} and \ref{fig:correlation-decay}. 
The meanings of the filled colors and line types are the same as in 
Figs.~\ref{fig:allowed-standard-param} and \ref{fig:correlation-decay}. 

As we can see from these plots that, in general, there are no 
significant differences among the allowed regions shown by 
the filled colors (no decay in input but allowed in the fit), 
solid black curve (standard oscillation fit without decay) 
and black dashed curves (decay effect both in the input and in the fit). 
Therefore, as a whole, the impact of the decay is rather small.
Furthermore, we note that there is no significant newly 
induced correlations due to the decay effect. 

%%%%%%%%%%%%%%%% FIG 5 %%%%%%%%%%%%%%%%%%
\begin{figure}[H]
\begin{center}
\vspace{-0.4cm}
\includegraphics[bb=0 0 614.88 1207.92,width=0.80\textwidth]{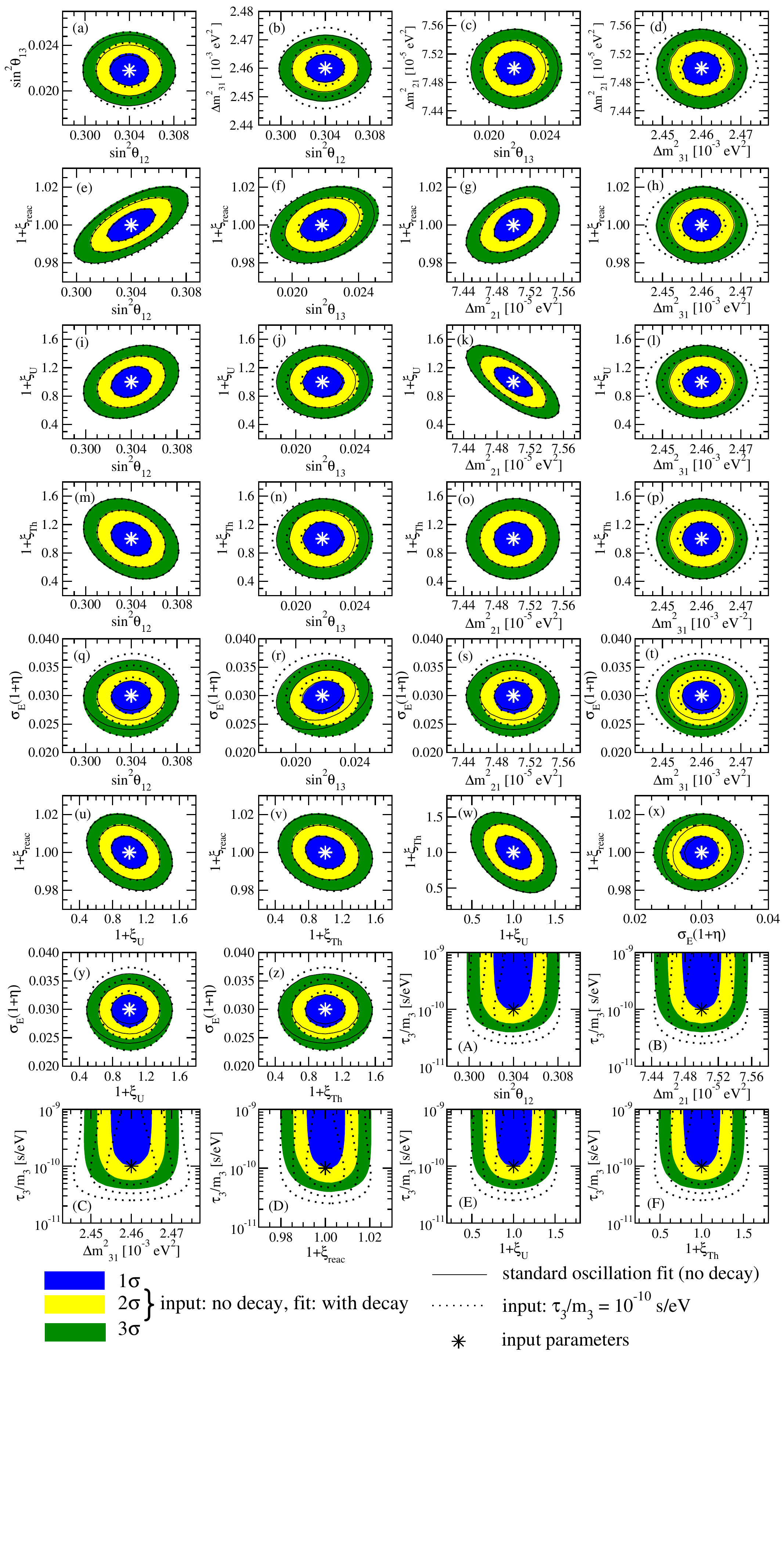}
\end{center}
\vspace{-3.6cm}
\caption{Allowed regions in the space spanned by all the combinations 
of parameters not shown in Sec.~\ref{sec:impact-decay} 
and in the previous subsection.
The filled color regions correspond to the case where no decay 
is considered for input but allowed in the fit. 
The black solid curves correspond to the case of the standard
oscillation fit without decay whereas the black dotted 
curves correspond to the case where the true value of 
$\tau_3/m_3$ is assumed to be $10^{-10}$ s/eV. 
}
\label{fig:allowed-regions-others}
\end{figure}
%%%%%%%%%%%%%%%% FIG 5 %%%%%%%%%%%%%%%%%%

%\newpage 

\begin{acknowledgments}
This work was supported by the Brazilian Funding Agencies, 
Funda\c{c}\~ao de Amparo \`a Pesquisa do Estado do Rio de Janeiro (FAPERJ), 
Conselho Nacional de Ci\^encia e Tecnologia (CNPq) and Coordena\c{c}\~ao 
de Aperfei\c{c}oamento de Pessoal de N\'{\i}vel Superior (CAPES).
We thank Anatael Cabrera for stimulating discussions and 
valuable suggestions.
H.M. thanks Universidade de S\~ao Paulo for the great opportunity of 
stay under ``Programa de Bolsas para Professors Visitantes
Internacionais na USP''.  
His visit to Instituto de F\'{\i}sica Te\'orica, Universidad Aut\'onoma 
de Madrid, where the last part of this work is completed, was supported 
by the Spanish MINECO's ``Centro de Excelencia Severo Ochoa'' 
Programme under grant SEV-2012-0249.
H.N. thanks the organizers of the INT Program INT-15-2a
``Neutrino Astrophysics and Fundamental Properties'' 
held at the Institute for Nuclear Theory, University of Washington,
where this work was completed, 
for the invitation to participate in the program.
H.M. and H.N. thank the Kavli Institute for Theoretical Physics in 
UC Santa Barbara for its hospitality, where part of this work was
done: This research was supported in part by the National 
Science Foundation under Grant No. NSF PHY11-25915. 

\end{acknowledgments}

\end{document}